\documentclass[twocolumn,prd,showpacs,superscriptaddress,groupedaddress,preprintnumbers,nofootinbib]{revtex4-1}
\pdfoutput=1
\usepackage{graphicx,amsfonts,amsmath,amssymb,epsfig,color, mathrsfs,latexsym,url,wasysym,float,xfrac}
\newcommand{\be}{\begin{equation}}
\newcommand{\ee}{\end{equation}}
\newcommand{\nn}{\nonumber}
\newcommand{\ba}{\begin{eqnarray}}
\newcommand{\ea}{\end{eqnarray}}
\newcommand{\mpl}{m_{\rm Pl}}
\newcommand{\HL}{H_\Lambda}
\newcommand{\Hi}{H_{\rm inf}}
\newcommand{\Heq}{H_{\rm eq}}

\newcommand{\Hrad}{H_{\rm rad}}
\newcommand{\Trad}{T_{\rm rad}}
\newcommand{\TBBN}{T_{\rm BBN}}

\newcommand{\ls}{\lesssim}
\newcommand{\gs}{\gtrsim}

\begin{document}

\title{Non-Standard Cosmological Models and the trans-Planckian Censorship Conjecture}

\author{Mahdi Torabian$^*$}
\affiliation{Department of Physics, Sharif University of Technology, Azadi Ave, 11155-9161, Tehran, Iran} 

\begin{abstract}
The trans-Planckian censorship conjecture (TCC) puts an upper bound on the life-time of de Sitter spacetimes. It has immediate consequences for inflationary cosmology. In the standard paradigm, the universe has experienced a single stage of inflation and follows a thermal history. Then, the TCC puts an upper bound on the Hubble parameter during inflation $H_{\rm inf}$ around $0.1$ GeV. Consequently, it implies a severe fine-tuning in initial condition for inflation and non-detection of primordial gravitational waves. In this note, we study non-standard cosmological paradigms with non-thermal history and/or multiple stages of inflations. It is motivated by string theory compactifications and axiverse scenarios in which the modulus/axion fields are effective in the early universe. In early matter domination the TCC bound on $H_{\rm inf}$ can be raised up to 3 orders of magnitude. In multiple inflationary scenarios the upper bound on the observable inflation can be raised up $10^{14}$ GeV to touch the {\it Planck 2018}  bound.
\end{abstract}

\preprint{SUT/Physics-nnn}
\maketitle

\subsection*{Introduction} 
The swampland program offers a list of criteria that every consistent theory of quantum gravity must admit \cite{Obied:2018sgi,Ooguri:2018wrx,Agrawal:2018own,Garg:2018reu} (see \cite{Brennan:2017rbf,Palti:2019pca} for review). The conditions are deduced from string theory constructions and it is believed that they must be proven based on quantum gravity principles. 
Recently, another swampland condition, the trans-Planckian censorship conjecture (TCC), is proposed so that sub-Planckian quantum fluctuations never get classicalized and must remain quantum \cite{Bedroya:2019snp}. It fits into the common lore that in order to explain physics in the IR, we do not need to know anything about deep UV. Equivalently, it states that the life time of a de Sitter (dS) vacuum is bounded. A dS vacuum is accompanied by either a Minkowski or an anti dS or a deeper dS vacua.  

The TCC has an immediate consequence in accelerating expanding spacetimes with shrinking comoving Hubble radius {\it a.k.a} inflationary backgrounds. This is particularly fascinating as inflation has become the dominant paradigm for the early universe to set the initial conditions of hot big-bang cosmology. On this backgrounds, the wavelength of quantum modes can be stretched beyond the horizon and become classical (see \cite{Baumann:2009ds} for a review). Simple models of inflation predict adiabatic, nearly scale-invariant and  Gaussian perturbations. It explains the observed anisotropies in the CMB radiation and seeds the large scale structures in the late universe.   The TCC then implies that there cannot be enough time of inflationary phase so that quantum fluctuations with wavelength smaller than the Planck length become superhorizon and non-dynamical. If it happens in a model, then that belongs to the swampland. It offers a solution to the trans-Planckian problem \cite{Martin:2000xs,Brandenberger:2000wr,Brandenberger:2012aj,Kaloper:2002cs} as it never happens in a field theory consistent with quantum gravity. 

Assuming, for simplicity, that the Hubble scale during inflation is constant, then the TCC implies
 \be\label{TCC} e^N <\frac{\mpl}{\Hi}.\ee     
Given a Hubble scale $\Hi$, it sets an upper bound on the number of {\it e}-folds $N$ and vice versa. On the other hand, inflation needs to last long enough so that the comoving Hubble radius shrinks to a smooth patch to explain the present (almost homogeneous) horizon. It demands some minimum number of {\it e}-folds which is determined by its succeeding cosmic evolution. Assuming radiation domination from the end of inflation to the time radiation-matter equality (the standard paradigm), the TCC offers an upper bound on the Hubble rate during inflation $\Hi\ls0.1$ GeV \cite{Bedroya:2019tba}. It suggests that the amplitude of primordial gravitation waves is negligible and the scalar to tensor perturbation ratio is $r<10^{-30}$. Moreover, it implies that the initial condition of inflation is extremely fine-tuned as $\epsilon<10^{-31}$, especially that the swampland distant conjecture restricts field excursion and the conventional solutions to the initial condition problem cannot be applied.  

The above upper bounds can be modified if the universe follows a non-standard cosmological time-line (related studies see recent papers \cite{Cai:2019hge,Tenkanen:2019wsd,Das:2019hto,Mizuno:2019bxy,Brahma:2019unn,Dhuria:2019oyf}). There could be epochs of matter domination and subsequent radiation domination before the onset of nucleosynthesis. The minimum number of $e$-folds is slightly sensitive to these details. Moreover, there could be multiple stages of inflation each gives some {\it e}-folds to the scale factor. The total number of {\it e}-folds resolve the horizon problem. There is an observable inflations that explains the slight inhomogeneities and it is followed by possibly many other. The TCC applies to each period of inflation to prevent sub-Planckian modes exit the horizon. 

The non-standard scenarios are in fact motivated by string theory compactifications and its axiverse. They generically predict the existence of light (pseudo-)scalar modulus (axion) fields with gravitational couplings to the Standard Model sector. These fields are effective in the early universe and could change its expansion rate (see \cite{Kachru:2003aw,Balasubramanian:2005zx,Conlon:2005ki,Acharya:2007rc,Kane:2015jia,DiMarco:2018bnw} for moduli and \cite{Arvanitaki:2009fg,Marsh:2011gr,Cicoli:2012sz,Kamionkowski:2014zda,Kawasaki:2010ux} for axion dynamics).

In this paper, we study these non-standard scenarios in the light of the TCC and we find that indeed the Hubble scale during inflation could be raised. First, we study a single inflation model with subsequent non-standard fluid domination and then we analyze multiple inflationary paradigms.

\subsection*{General TCC bound in non-standard cosmologies}
We start with a general study of cosmological evolution. We assume that the observable (first) stage of inflation starts at $t_{\rm ini}$ and ends at $t_{\rm end}$. It follows by an era dominated with a fluid with equation of state (EoS) parameter $w$. 
%It could be matter, radiation or some exotic dominated. 
Then it follows by subsequent eras marked by time $t_i$ with EoS parameter $w_i$. It could be dominated by any fluid including dark energy {\it i.e.} secondary stages of inflation. Finally, there is the time of last radiation dominated era, marked by $t_{\rm rad}$ which can be any time before the nucleosynthesis. Thus, $\Trad\gs\TBBN\sim 10$ MeV and by definition there is no entropy injection afterwards. It marks the dawn of standard hot big bang cosmology. For simplicity, we assume instantaneous transition between each two epochs. The time-line is shown in figure 1.

\vspace*{-5mm}\begin{center}\begin{figure}[h!]\includegraphics[scale=.5]{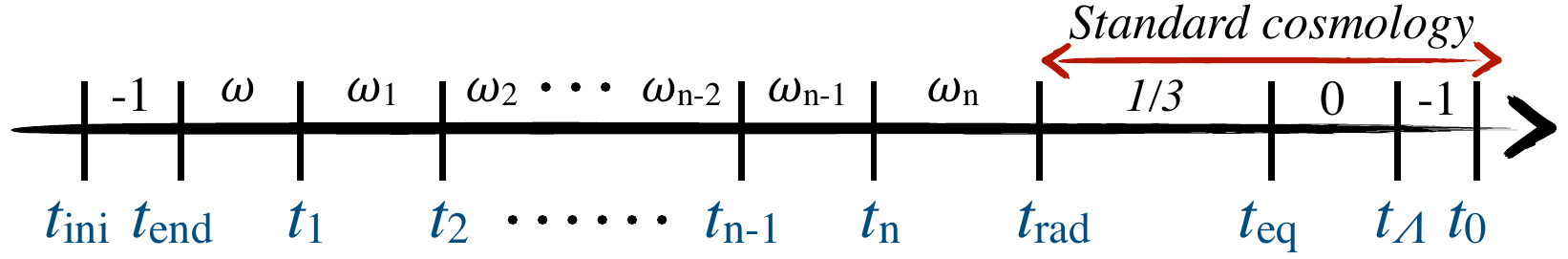}\caption{Time-line of the universe indicating eras with EoS.}\end{figure}\end{center} \vspace*{-8mm}      

Depending on the EoS parameter of the era, the comoving Hubble radius $(Ha)^{-1}$ is either contracting (for $w>-\sfrac{1}{3}$) or expanding (for $w<-\sfrac{1}{3}$) or constant (for $w=-\sfrac{1}{3}$). The TCC assures that non of the inflating phases lasts too long.  To explain near homogeneity at large scale, there must be at least one epoch with shrinking comoving Hubble scale so that the largest present observable scale was inside the Hubble horizon during inflation $k_\Lambda^{-1}=(a_\Lambda \HL)^{-1}\leq(a_{\rm ini} \Hi)^{-1}$ . Thus \cite{Liddle:2003as}
\ba\label{N} 1&\geq& \frac{a_{\rm ini}\Hi}{a_\Lambda \HL} \cr &=&\!\frac{\Hi}{\HL}e^{-N}\!\Big[\frac{a_{\rm end}}{a_1}\frac{a_1}{a_2}\cdots\frac{a_{n-1}}{a_n}\frac{a_n}{a_{\rm rad}}\Big]\!\frac{a_{\rm rad}}%{a_{\rm eq}}\frac{a_{\rm eq}}{a_\Lambda}\frac{a_\Lambda}
{a_0}(1+z_\Lambda),\quad\ \ea
where $\HL\approx H_0$ is the present Hubble rate, $z_\Lambda>0$ is the redshift factor from the c.c. domination and $N =\ln(a_{\rm end}/a_{\rm ini})$ is the number of {\it e}-folds. For the sake of clarification explicitly write scale factors of the intermediate stages. Scale factors can be substituted by the Hubble scales at the beginning of each epoch
\be \frac{a_i}{a_{i+1}}=\left(\frac{H_{i+1}}{H_i}\right)^{\sfrac{2}{3(1+w_i)}}\ {\rm or}\quad \frac{a_i}{a_{i+1}}=e^{-N_i}, \ee
for $w_i\neq-1$ and $w_i=-1$ respectively. Moreover,
\be %\frac{a_{\rm rad}}{a_{\rm eq}}\frac{a_{\rm eq}}{a_\Lambda}\frac{a_\Lambda}{a_0}
\frac{a_{\rm rad}}{a_0}=\HL^{\sfrac{2}{3}}\Heq^{\sfrac{-1}{6}}\Hrad^{\sfrac{-1}{2}}%e^{-N_\Lambda}
(1+z_\Lambda)\sim \frac{T_0}{T_{\rm rad}}. \ee
Throughout this paper, we drop powers of $g_*$ which is the number of relativistic species. Moreover, we use $H_0=2.13h\times 10^{-33}\ {\rm eV}$ with $h=0.6766\pm0.0042$, $T_0=6.626\times 10^{-4}$ eV \cite{Aghanim:2018eyx} and assume $w\approx -1$ during inflation. 

{\it Single Inflation} In eq. \eqref{N} we singled out one inflationary stage as it is the minimum number the universe needs to get to evolve to its present state. We use \eqref{TCC} and \eqref{N} to put an upper bound on the Hubble scale of inflation that explains near homogeneity
\ba\label{master} \Hi^{2-\sfrac{2}{3(1+w)}}\!\!<\! \mpl\HL T_0^{-1}\Trad H_1^{-\sfrac{2}{3(1+w)}}\frac{a_2}{a_1}%\frac{a_{n-2}}{a_{n-1}}
\!\cdots\!\frac{a_n}{a_{n-1}}\!\frac{a_{\rm rad}}{a_n}.\nn\\ \ea
The above equation indicates that the smaller the EoS parameter after inflation, the bigger the upper bound on $\Hi$. In the following, we study different cosmological models which are illustrated in figure 2.
\vspace*{0mm}\begin{center}\begin{figure}[b!]\includegraphics[scale=.39]{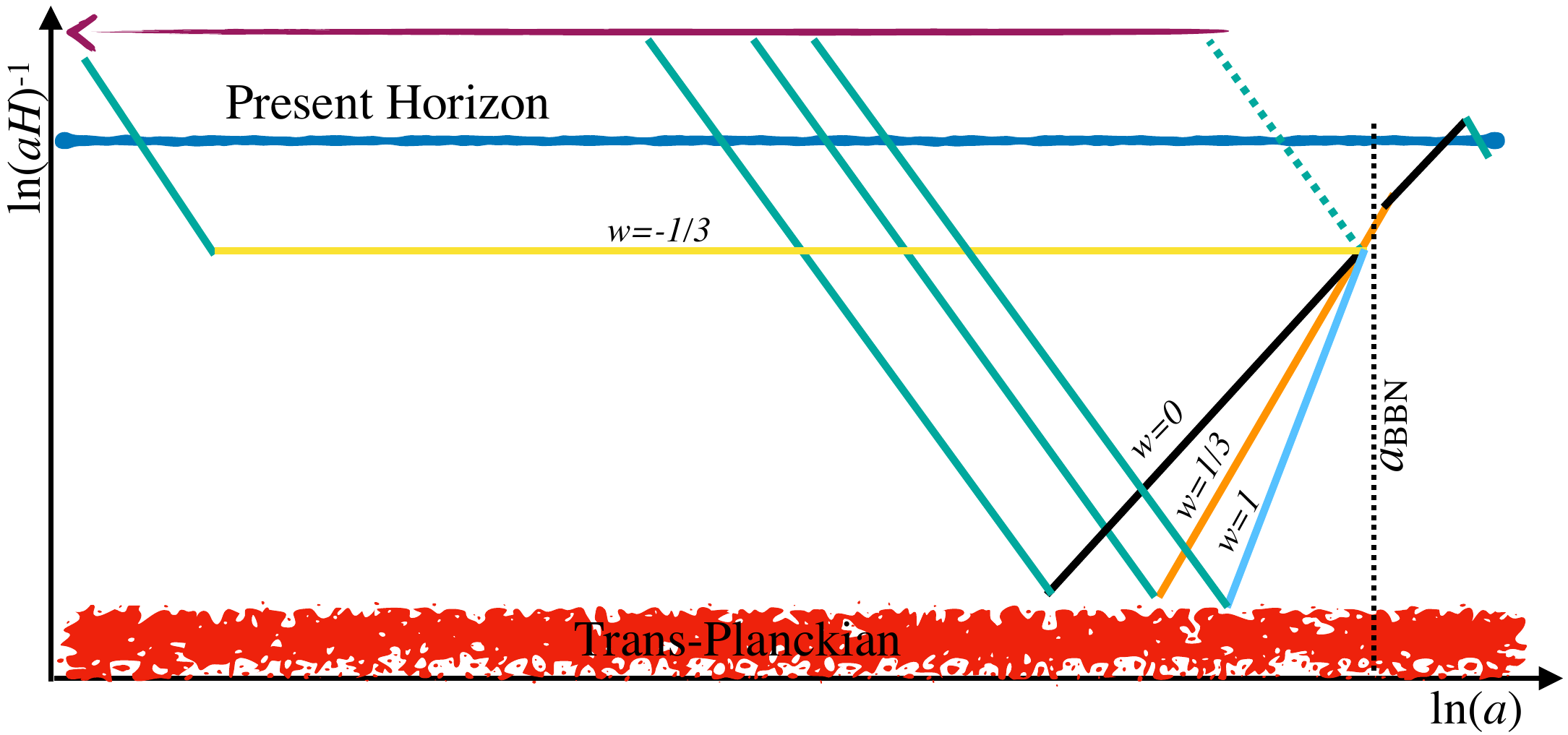}\caption{Single inflation followed by different post-inflationary epochs. Left-headed arrow shows the direction of increasing Hubble rate. Cyan lines show inflationary phases, with highest possible Hubble scale, followed by different cosmologies. The dotted line shows the lowest possible scale inflation.}\end{figure}\end{center} \vspace*{0mm}   

\vspace*{-8mm}{\it I.} In the simplest scenario, the universe is radiation dominated instantly after the end of inflation until the time of equality thus $w=\sfrac{1}{3}$.
%Thus, we put $w=\sfrac{1}{3}$ and $t_1=t_{\rm rad}$ and $H_1=\Hrad$ in \eqref{master} and
Then \eqref{master} implies
\ba \Hi^{\sfrac{3}{2}}< \mpl \HL T_0^{-1} \Trad \Hrad^{-\sfrac{1}{2}} \sim \mpl^{\sfrac{3}{2}} \HL T_0^{-1},\ \ea
and so $\Hi\ls0.1$ GeV as first reported in \cite{Bedroya:2019tba}. 

{\it II.} If the dynamics of the universe is dominated with some exotic matter by $w\approx-\sfrac{1}{3}$ (like cosmic string network \cite{Burgess:2005sb}
) after the end of inflation until the time of nucleosynthesis \cite{Mizuno:2019bxy}, then, \eqref{master} gives
\be \Hi< \mpl^2\HL T_0^{-1}\Trad^{-1},\ee
which implies  $\Hi\ls 10^{14} {\rm GeV}\cdot(10{\rm MeV}/\Trad)$. 
%This almost touches the Planck 2018 bound from gravitational waves.

{\it III.} For the case of post-inflationary kinetic domination with $w=1$ up to nucleosynthesis we find
\ba \Hi^{\sfrac{5}{3}}< \mpl\HL T_0^{-1}\Trad\Hrad^{\sfrac{-1}{3}}\sim\mpl^{\sfrac{4}{3}}\HL T_0^{-1}\Trad^{\sfrac{1}{3}},\ \ \ea
which implies $\Hi<$ MeV$\cdot(10{\rm MeV}/\Trad)^{\sfrac{1}{5}}$.  

{\it IV.} In a different cosmic evolution, we assume that the universe is matter dominated after inflation until the time of nucleosynthesis so $w=0$. Then, it gets reheated and radiation takes over to commence synthesis of light isotopes. In this case, \eqref{master} implies
\ba \Hi^{\sfrac{4}{3}}< \mpl\HL T_0^{-1}\Trad\Hrad^{\sfrac{-2}{3}}\sim\mpl^{\sfrac{5}{3}}\HL T_0^{-1}\Trad^{\sfrac{-1}{3}}.\ \ \ea
We find that $\Hi\ls100$ GeV$\cdot(10{\rm MeV}/\Trad)^{\sfrac{1}{4}}$ and the energy scale of inflation is  around $10^{10}$ GeV.
 
As a special case and for completeness, here we study an early brief matter domination before nucleosynthesis {\it a.k.a.} moduli cosmology . This scenario is in particular motivated by string theory constructions. 
String compactifications predict moduli (axion) fields with no direct coupling to the visible sector. The flat directions of these scalars are lifted after supersymmetry breaking. They are rather light and  during inflation they are normally misaligned and stay far from their zero values.
When the Hubble rate is about a modulus mass $H\sim m_\varphi$, it commences coherent oscillation about its minimum. The scalar condensates, as a dust \cite{Turner:1983he}, quickly take over radiation domination. The universe enters into a matter domination epoch until the Hubble scale becomes comparable to the modulus decay rate $H\sim \Gamma_\varphi$. The decay products thermalize and reheat the universe. In order to kick nucleosynthesis the reheat temperature must be greater than few 10 MeV. For gravitationally coupled moduli fields $\Gamma_\varphi\sim m^3_\varphi/\mpl^2$
and thus there is a lower bound on the lightest modulus mass $m_\varphi\gs10$ TeV. We set $H_1=m_\varphi$ in \eqref{master} and we find
\ba \Hi^{\sfrac{3}{2}}< &&\ \mpl\HL T_0^{-1}\Trad m_\varphi^{-\sfrac{1}{2}}(m_\varphi/\Gamma_\varphi)^{\sfrac{2}{3}}\cr &&\sim \mpl^{\sfrac{11}{6}}\HL T_0^{-1}m_\varphi^{-\sfrac{1}{3}}.\ea
Given the bound on the modulus mass, we find the Hubble rate during inflation can be enhanced at most by 3 orders of magnitude to $100$ GeV.

In a generic construction with many moduli fields, if there is no unnatural mass splitting, then the universe is matter dominated from the mass scale of the heavier modulus $H\sim m_{\varphi_h}$ until the decay rate of the lightest one $H\sim \Gamma_{\varphi_l}$ \cite{Acharya:2019pas}. 
%For  $H_1=m_{\varphi_h}$  and $\Hrad=\Gamma_{\varphi_l}$ in
Then \eqref{master} implies
\ba \Hi^{\sfrac{3}{2}}< &&\ \mpl\HL T_0^{-1}\Trad m_{\varphi_h}^{-\sfrac{1}{2}}(m_{\varphi_h}/\Gamma_{\varphi_l})^{\sfrac{2}{3}}\cr &&\sim \mpl^{\sfrac{11}{6}}\HL T_0^{-1}m_{\varphi_l}^{-\sfrac{1}{2}}m_{\varphi_h}^{\sfrac{1}{6}}. \ea
Moreover, if inflation is followed by its matter domination which subsequently decays at $H\sim \Gamma_\phi$, then %applying $H_1=\Gamma_\phi$, $\Hrad=\Gamma_{\varphi}$ and $H_{\rm osc}=m_\varphi$ in 
\eqref{master} gives
\ba \Hi^{\sfrac{4}{3}}< &&\ \mpl\HL T_0^{-1}\Trad \Gamma_\phi^{-\sfrac{2}{3}} (\Gamma_\phi/m_\varphi)^{\sfrac{1}{2}}(m_\varphi/\Gamma_\varphi)^{\sfrac{2}{3}}\cr &&\sim \mpl^{\sfrac{11}{6}}\HL T_0^{-1}\Gamma_\phi^{\sfrac{1}{6}}m_\varphi^{-\sfrac{1}{3}}.\ea
We conclude that in moduli cosmology $\Hi$ is enhanced at most by 3 orders of magnitude. 

{\it Multiple Inflation} Here we study a cosmological model with multiple inflationary stages and demand that all admit to the TCC.  Secondary inflations of low scale follow the observable high scale one and make the universe large and old enough. Collectively, they explain the present horizon and its little inhomogeneities. We assume that the scale entering the horizon today has left the horizon during the first inflationary epoch. Namely, the first epoch of inflation generates perturbations that we observe today as anisotropies in the CMB. In fact, the current CMB observations probe up to 6$\sim$7 {\it e}-folds after the present scale exited the horizon. Thus we expect that the first inflation lasted for plus 6 {\it e}-folds. We preserve those 6$\sim$7 {\it e}-folds from any subsequent dynamics by demanding scales $k$ up to 3 orders of magnitude larger than the present scale never become subhorizon until today. Therefore, we constrain intermediate stages with expanding comoving Hubble horizon so that the CMB modes do not get to re-enter the horizon and exit again during succeeding inflation (see figure 3 for illustration). 

\vspace*{-4mm}\begin{center}\begin{figure}[t!]\includegraphics[scale=.5]{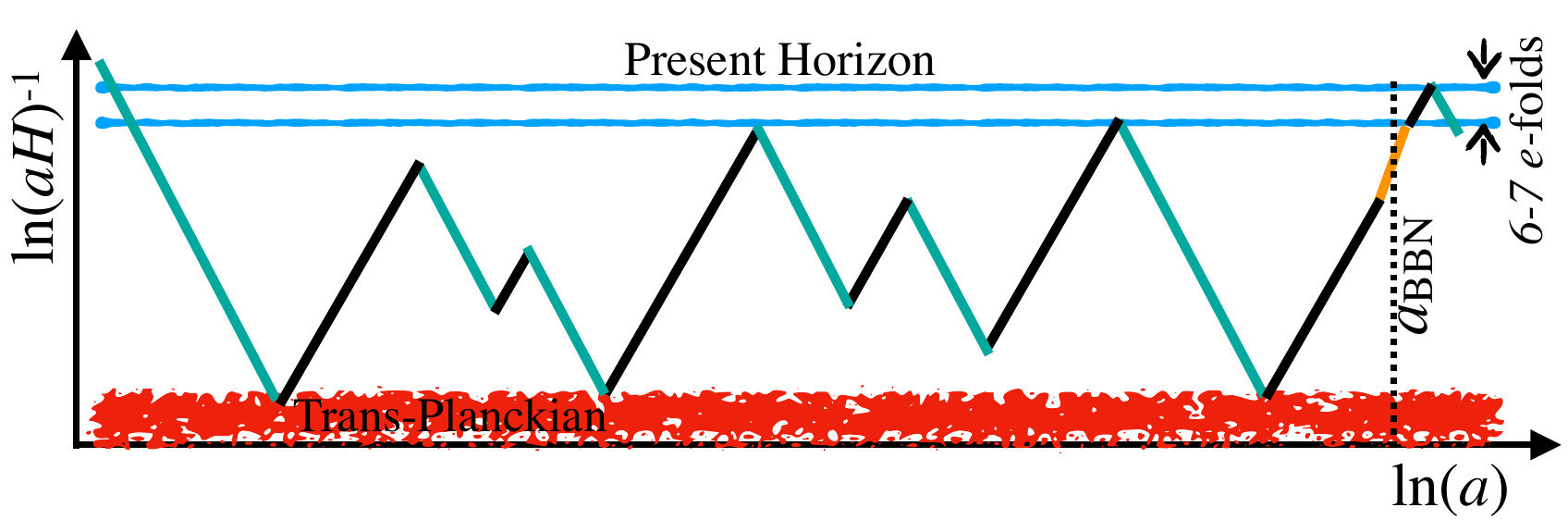}\caption{A multiple inflationary scenario for the evolution of the universe. The observable high scale inflation is succeeded by many low scale ones. All admit to the TCC. The observable modes on the CMB are protected from re-entry and exiting again.}\end{figure}\end{center} \vspace*{-0mm}      

Applying  \eqref{master} we compute the upper bound on the Hubble rate of the observable $\Hi$ in  multiple inflationary paradigm as
\ba\label{master-2} \Hi^{2-\sfrac{2}{3(1+w)}}<&&\ \mpl\HL T_0^{-1}\Trad  e^{\sum_iN_i}\cr &&\ \times H_1^{-\sfrac{2}{3(1+w)}}\prod_i\Big(\frac{H_j}{H_{j+1}}\Big)^{\sfrac{2}{3(1+w_j)}}.\ \ea
We understand that the total number of {\it e}-folds is $N_e=N+\sum_{i=1}N_i$ and we need a minimum of them to accommodate near homogeneity. In the following we study two simple scenarios with multiple inflations.

{\it I.} We consider a model in which each inflation is followed by instant reheating and thus there is no matter domination until the equality time. Then \eqref{master-2} implies
\ba \Hi^{\sfrac{3}{2}}\ls&& \mpl\HL T_0^{-1}\Trad H_1^{-\sfrac{1}{2}}  e^{\sum_iN_i}\prod_{i=1}\Big(\frac{H_i}{H_{i+1}}\Big)^{\sfrac{1}{2}}\cr
\sim&&\ \mpl^{\sfrac{3}{2}} \HL T_0^{-1}e^{\sum_iN_i}.\ea
Clearly, we find that the original TCC bound on inflationary scale is pretty much relaxed
\be \Hi\ls %\mpl(\HL/T_0)^{\sfrac{2}{3}}e^{\frac{2}{3}\sum_i N_i}\sim
0.1 {\rm GeV}\times  e^{\frac{2}{3}\sum_i N_i}.\ee

{\it II.} We look at a scenario with matter domination between inflationary stages with late radiation domination before nucleosynthesis. We find 
\ba \Hi^{\sfrac{4}{3}}\ls&& \mpl\HL T_0^{-1}\Trad H_1^{-\sfrac{2}{3}}e^{\sum_iN_i} \prod_{i=1}\Big(\frac{H_i}{H_{i+1}}\Big)^{\sfrac{2}{3}} \cr
\sim&&\ \mpl^{\sfrac{5}{3}} \HL T_0^{-1}\Trad^{-\sfrac{1}{3}}e^{\sum_iN_i},\ea
which implies 
\be \Hi\ls100 {\rm GeV}\times e^{{\sfrac{3}{4}}\sum_iN_i}\cdot(10{\rm MeV}/\Trad)^{\sfrac{1}{3}}.\ee
Evidently, multiple inflationary scenarios radically modify the upper bound on the observable inflation. It can be raised to the upper bound imposed by {\it Planck 2018} results, namely, $\Hi\ls10^{14}$ GeV \cite{Akrami:2018odb}.
%The best we can get with some exotic (curvature-like) matter
%\be \Hi\ls \mpl^2\HL T_0^{-1}\Trad^{-1}e^{\sum_iN_i},\ee
%which implies  $\Hi\ls 10^{14} {\rm GeV}\times e^{\sum_iN_i}\cdot(10{\rm MeV}/\Trad)$

\subsection*{Conclusion}
In this note, we revisited the TCC bound on the scale inflation in the context of non-standard cosmologies. As expected, we find that the smaller the EoS parameters the bigger the upper bound on the Hubble rate. In cosmological scenarios with early matter domination, motivated by string compactifications/axiverse, the Hubble scale during inflation can be raised up to 3 orders of magnitude. In cosmological models with secondary inflations, we find that the upper bound can be raised even to the {\it Planck 2018} limit. If primordial gravitational waves are detected in future observations, then it can be accommodated in multi-stage inflationary scenarios with one brief high scale inflation followed by many low scale ones. A detailed study of the scale and duration of successive inflations is postponed to a future work \cite{torabian}.

\paragraph*{Acknowledgments}
The author is thankful to Ali-Akbar Abolhasani for inspiring discussions on CMB observable modes.  This work is supported by the research deputy of SUT.
\ \\$^*$ Electronic address: mahdi.torabian@sharif.ir

\end{document}